\documentclass[twocolumn,aps,prl]{revtex4-1}

\usepackage[T1]{fontenc}
\usepackage{graphicx}
\usepackage{dcolumn}
\usepackage{bm}
\usepackage{geometry}
\usepackage{color}
\usepackage{siunitx}

\newcommand{\Th}{\mathrm{Th}}

\begin{document}

\preprint{APS/123-QED}

\title{Observation of an unexpected negative isotope shift in $^{229}$Th$^+$ and its theoretical explanation}
\author{M.V. Okhapkin, D.M. Meier,  E. Peik}

\email{Corresponding author E-mail address: ekkehard.peik@ptb.de}
\affiliation{Physikalisch-Technische Bundesanstalt, Bundesallee 100, 38116 Braunschweig, Germany
}
\author{M.S. Safronova$^{1,2}$, M.G. Kozlov$^{3,4}$, S.G. Porsev$^{1,3}$}
\affiliation{$^{1}$Department of Physics and Astronomy, University of Delaware, Newark, Delaware 19716, USA\\
$^{2}$Joint Quantum Institute, National Institute of Standards and Technology
and the University of Maryland, Gaithersburg, Maryland 20899, USA\\
$^{3}$Petersburg Nuclear Physics Institute, Gatchina 188300, Russia\\
$^{4}$St. Petersburg Electrotechnical University LETI, St. Petersburg 197376, Russia
}

\date{\today}%

\begin{abstract}
We have measured the hyperfine structure and isotope shifts of the \SI{402.0}{\nm} and \SI{399.6}{\nm} resonance lines in  $^{229}$Th$^+$. These transitions could provide pathways towards the excitation of the $^{229}$Th low-energy isomeric nuclear state. An unexpected negative isotope shift relative to $^{232}$Th$^+$ is observed for the \SI{399.6}{\nm} line, indicating a strong Coulomb coupling of the excited state to the nucleus. We have developed a new all-order approach to the isotope shift calculations that is generally applicable to heavy atoms and ions with several valence electrons. The theoretical calculations provide an explanation for the negative isotope shift of the \SI{399.6}{\nm} transition and yield a corrected classification of the excited state. The calculated isotope shifts are in good agreement with experimental values. 
\end{abstract}

\pacs{
      }

\maketitle

The existence of a low-energy nuclear isomeric transition in  $^{229}$Th \cite{Helmer:1994, Beck:2007, Beck:2009}  has stimulated the development of novel ideas and concepts in the borderland between atomic and nuclear physics \cite{Matinyan:1998, Tkalya:2003}, with conceivable applications like nuclear clocks \cite{Peik:2003, Campbell:2012, Rellergert:2010, Kazakov:2012, Peik:2015} and $\gamma$-ray lasers \cite{Tkalya:2011}. 
The wavelength of this very narrow nuclear transition is predicted to be 160(10)~nm ~\cite{Beck:2007, Beck:2009}. The big uncertainty in the transition wavelength, its location in the vacuum ultraviolet range,  and the radioactivity of $^{229}$Th with a halflife of about 7900 years present experimental difficulties that have so far prevented optical spectroscopy of the transition. A promising approach towards an excitation of the isomer is the use of electronic bridge or NEET processes (nuclear excitation by electron transition)\cite{Tkalya:1996, Karpeshin:1999, Porsev:2010}, that involves finding  sufficiently strong
 electronic transitions close to the nuclear resonance.  The Th$^+$ ion, with its large density of states, offers a high probability of locating such electronic bridge pathways. Using the long-lived $^{232}$Th isotope we have investigated two-photon 
excitation of electronic levels of Th$^+$ in the expected energy range of the isomer and have found 43 previously unknown levels \cite{Herrera:2013}.

A well known influence of the nucleus on electronic spectra of atoms is the isotope shift (IS) that for heavy atoms is dominated by differences in the overlap of nuclear and electronic charge distributions, the so called field shift (FS).
Extensive studies of isotope shifts in the dense level schemes of Th and Th$^+$ have been performed \cite{Stukenbroeker:1953,Vernyi:1959,Vernyi:1960,Engleman:1984} providing shifts of more than 350 excited states of  $^{230}$Th$^+$ with respect to $^{232}$Th$^+$ \cite{Engleman:1984}. In all but 3 cases, the shift of the excited state is smaller than that of the $6d^2 7s$ ground state (GS), indicating the smaller overlap of the excited electronic configuration with the nucleus for almost all states.  Although a nearly complete compilation of the 15 lowest electron configurations is published \cite{Zalubas:1974}, only very few low-energy excited states belonging to $fs^2$ and $s^2p$ configurations with strong Coulomb interaction with the nucleus are known in Th$^+$. Such states will be of relevance for the implementations of electronic bridge or NEET processes.

Based  on laser spectroscopy of trapped ions, we present here measurements of the isotope shifts and hyperfine structure (HFS)  of $^{229}\mathrm{Th}^{+}$ for two of the strongest Th$^+$ resonance lines at \SI{399.6}{\nm} and \SI{402.0}{\nm}.  
A positive IS of the $402~$nm  line of $^{229} \Th^{+}$ relative to $^{232} \Th^{+}$ of $19~$GHz has been reported already~\cite{Vernyi:1960}. Unexpectedly,  we observe a negative isotope shift (i.e. a lower transition frequency) of the $0_{3/2}\rightarrow 25027_{1/2}$ line at \SI{399.6}{\nm} with respect to $^{232}\mathrm{Th}^{+}$,
indicating a strong field shift of the excited state (states are labeled by their energy in cm$^{-1}$ and the angular momentum $J$). Since the $25027_{1/2}$ state was previously identified as  $5f6d^2$  \cite{Zalubas:1974},  a negative isotope shift would contradict the available IS data for Th$^+$ as discussed above. 

Resolving this issue requires precision theoretical IS calculations.
This is a  difficult endeavor due to several factors: (1) Th$^+$ is a heavy
trivalent system with electronic configurations containing $5f$ electrons that are very strongly correlated with
core electrons; (2) there is a very large number of energy levels between the ground state and the
$25027_{1/2}$ level (it is the 82nd odd level); (3) the state of interest is very strongly mixed with nearby states
complicating the calculations and identification of the levels.  Therefore, a very accurate
new approach is needed for the Th$^+$ IS calculation.

We have developed a method to calculate IS for multivalent heavy systems based on the hybrid approach that combines configuration interaction and all-order linearized coupled cluster methods (CI+all-order). While the CI+all-order method has been applied to calculating energies, transitions properties, and polarizabilities \cite{Safronova:2009,Safronova:2014,Safronova:2014a}, this is its first implementation for IS calculations.
This is also a first application of the CI+all-order method for such highly-excited states of multi-valent atoms for any system.
An ability to accurately calculate IS for heavy ions
 opens a pathway for the accurate  determination of nuclear radius differences between isotopes of heavy elements based on
\textit{ab initio} calculations.

In our experiments we use a linear Paul trap, described in Refs.~\cite{Herrera:2013,Herrera:2012}. The trap is loaded  with $\geq 10^5$ Th$^+$ ions by laser ablation from a dried $\Th(\mathrm{NO}_{3})_{4}$ solution deposited on a tungsten substrate. 
Collisions with argon buffer gas at \SI{0.1}{\Pa} pressure cool the ions to room 
temperature and quench the population of metastable states that are optically pumped 
by the laser excitation.  Pulsed excitation of the $24874_{5/2}$ and $25027_{1/2}$ 
states is provided by a CW extended-cavity diode laser (ECDL) with an output power of 
\SI{25}{\mW} passed through a fast acousto-optical modulator. 
The excitation pulse duration is \SI{70}{\ns} and the pulse repetition rate of \SI{1}
{\kHz} is adapted to the collisional quenching 
rate of about $\approx$\SI{1500}{/s} ~\cite{Herrera:2012}.  Fluorescence of the 
excited $\Th^{+}$ ions is detected using a photomultiplier (PMT). 
A fast gated integrator is used to evaluate the PMT signal during a detection 
window with a duration of about \SI{75}{\ns}. The signal is then integrated over 
several hundred pulses. A small fraction of the CW radiation of the excitation ECDL  is coupled to a Fabry - Perot interferometer used as a reference to measure frequency intervals.
After loading of Th$^+$ we observe simultaneously the fluorescence signal from both $^{229} \mathrm{Th}$ and $^{232} \mathrm{Th}$, indicating approximately equal abundances of both isotopes on the substrate.  When it is necessary to avoid the influence of $^{232} \Th^{+}$ ions we apply an isotope selective resonantly enhanced three-photon ionization to Th$^{2+}$ ~\cite{Herrera:2013} and eject these ions from the trap.

\begin{figure}[h]
\centering
\includegraphics[width=0.35\textwidth]{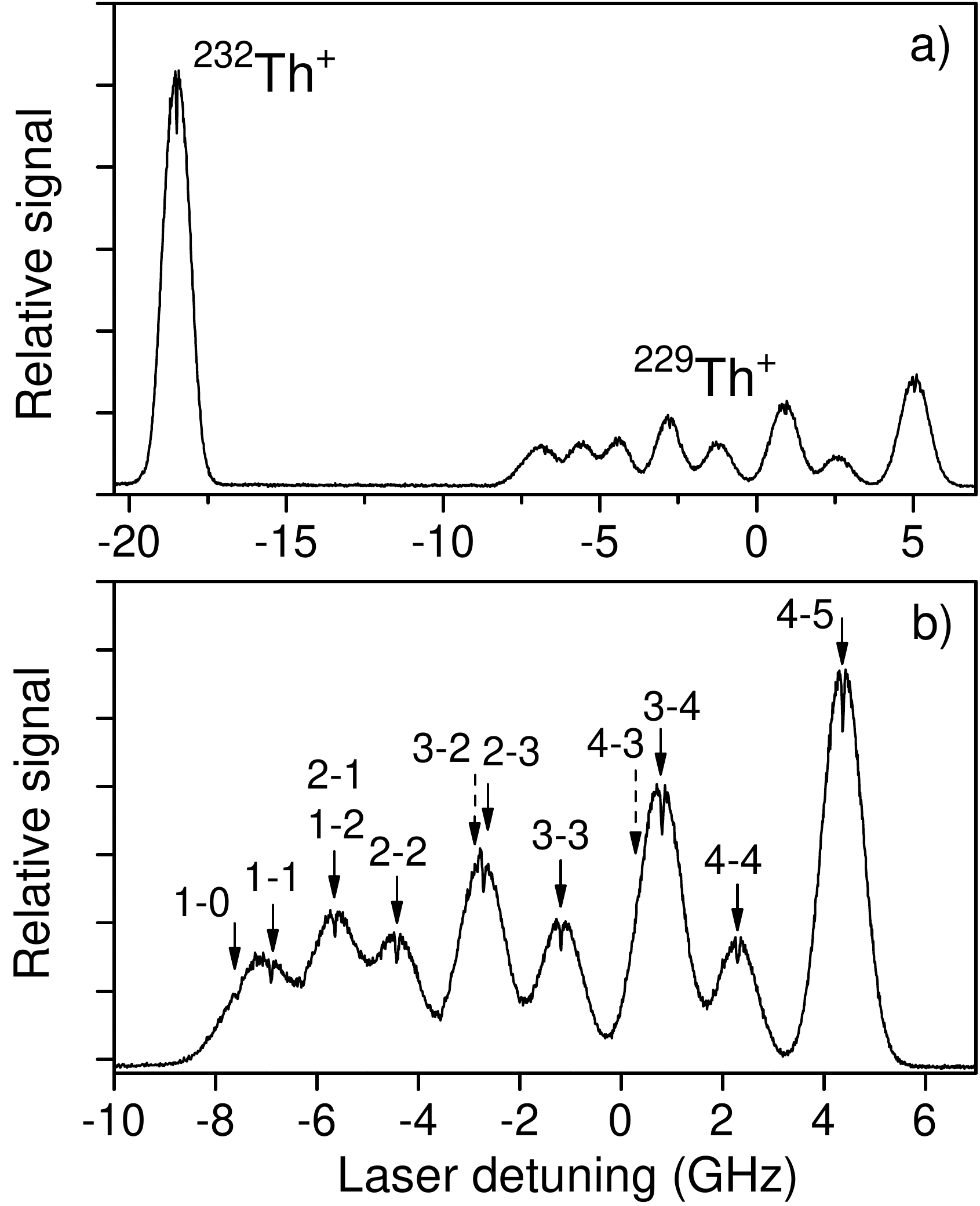}
\caption{\label{fig:hfs} a) Fluorescence signal of the \SI{402.0}{\nm} line of $^{229}\Th^{+}$ and $^{232}\Th^{+}$ isotopes. b) HFS signal of the \SI{402.0}{\nm} $^{229}\Th^{+}$ line. The detuning counts from the recalculated HFS centroid position of the $^{229}\Th^{+}$ isotope line.  The components are labeled with the total angular momenta of lower and upper hyperfine levels. The dashed arrows in (b) show the recalculated positions of the $3\rightarrow{2}$ and $4\rightarrow{3}$ resonances.}
\end{figure}

Determining the IS for  $^{229}\mathrm{Th}$ with a nuclear spin $I=5/2$ requires knowledge of the HFS centroid frequency. The HFS of the $0_{3/2}\rightarrow24874_{5/2}$  transition of $^{229}\mathrm{Th}^+$ at \SI{402.0}{\nm} consists of 12 lines with 4 components in each of the $\Delta F=-1,0,+1$ branches, where $F$ is the total angular momentum. We investigate the HFS  by Doppler-free saturation spectroscopy and 
the excitation spectrum of both  $^{229}\Th^{+}$ and $^{232}\Th^{+}$ isotopes is shown in Fig.~\ref{fig:hfs}a. Fig.~\ref{fig:hfs}b shows the HFS of $^{229}\Th^{+}$ with high resolution. The width (FWHM) of the resonances is $\geq$\SI{70}{\MHz} and the width of the Doppler broadened lines is $\approx$\SI{760}{\MHz}.
In Fig.~\ref{fig:hfs}b nine out of twelve possible resonances are resolved. The resonances caused by $F=1\rightarrow{F'=2}$ and $2\rightarrow{1}$ transitions are overlapping within the width of the Lamb dips. The deformation of the Doppler shape caused by component $3\rightarrow{2}$ is also partially observable on the graph.  The weak HFS component $4\rightarrow{3}$ is not observed due to the limited signal-to-noise ratio. 

\begin{figure}[h]
\centering
\includegraphics[width=0.36\textwidth]{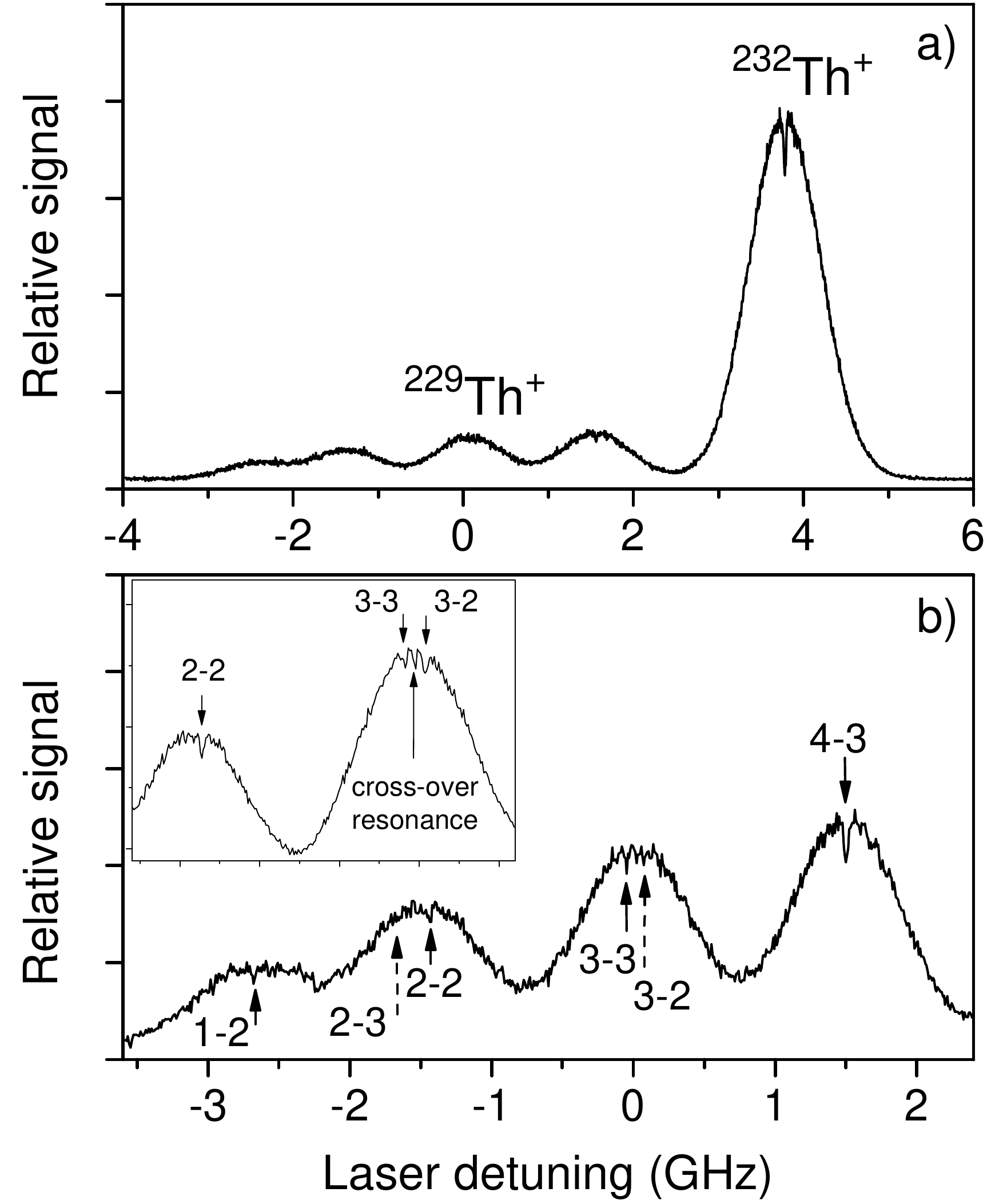}
\caption{\label{fig:hfs_399} a) Fluorescence signal of the \SI{399.0}{\nm} line of $^{229}\Th^{+}$ and $^{232}\Th^{+}$ isotopes. b) HFS signal of the \SI{399.6}{\nm} $^{229}\Th^{+}$ line. The detuning counts from the recalculated HFS centroid position of the $^{229}\Th^{+}$ isotope line, and the dashed arrows show the recalculated positions of the resonances $2\rightarrow{3}$ and $3\rightarrow{2}$.  The inset shows resolved Doppler-free resonances of the components $3\rightarrow{3}$ and $3\rightarrow{2}$ and the $\mathrm{V}$-type cross-over resonance.}
\end{figure}

The HFS of the \SI{399.6}{\nm} line  $0_{3/2}\rightarrow 25027_{1/2}$ is shown in Fig.~\ref{fig:hfs_399}a. The $25027_{1/2}$ state of $^{229}$Th$^+$ consists of two components with $F=2, 3$, therefore the HFS of the transition comprises 6 lines.
The hyperfine components of $^{229} \Th^{+}$ appear at frequencies below the position of the $^{232} \Th^{+}$ line. To exclude a possible masking of a hyperfine component with the $^{232} \Th^{+}$ Doppler broadened line we ejected $^{232}\Th^{+}$ with the technique mentioned above. This experiment shows that there is no hyperfine component of the $^{229} \Th^{+}$ overlapping with the $^{232}\Th^{+}$ line.  The saturation resonance of the transition $2\rightarrow{3}$ has a very weak intensity and is not observed in the experiment. The resonances caused by transitions $3\rightarrow{3}$ and $3\rightarrow{2}$ are resolved by scanning with longer integration time (see the inset in Fig. 2b). The dip in the middle correspond to a cross-over resonance of $\mathrm{V}$-type with the common ground state level $F=3~$ (see Fig.~\ref{fig:hfs_399}b).

In order to determine the hyperfine splitting factors $A$ and $B$ of the $\SI{402.0}{\nm}$ and $\SI{399.6}{\nm}$ lines we measure the frequency intervals between the Lamb dips relative to the transmission peaks of the interferometer. For the $\SI{402.0}{\nm}$ line the measured frequency intervals are fitted in a least squares adjustment to determine the $A$ and $B$ factors for the ground and the $24874_{5/2}$ states.
The algorithm attempts fits for all viable assignment combinations of $F$ values for the nine resolved transitions shown in Fig.~\ref{fig:hfs}b.
The result is shown in Table ~\ref{table:splittingrates}. 
The uncertainty results from inaccuracy of the measured frequency intervals, including a small contribution from the determination of the reference interferometer FSR. 
The HFS factors for the ground electronic state (GS) in $^{229}\mathrm{Th}^+$ are already known with high accuracy ~\cite{Kalber:1989} (see Table ~\ref{table:splittingrates}) and our results are in good agreement. 

For the $25027_{1/2}$ state the $B$ factor vanishes because of the absence of a quadrupole moment for $J=1/2$.
To determine the $A$ factor of this level we fix the $A$ and $B$ values for the GS. We use the same algorithm to fit the frequency intervals between resolved transitions of this line.
 The uncertainty of the $A$ value of the  $25027_{1/2}$  state is determined by the accuracy of the frequency intervals and the uncertainty of the GS factors. Due to the weaker signal-to-noise ratio and smaller $A$ factor of the excited state, the uncertainty of the fit is higher than that for the  $24874_{5/2}$ state. Using the obtained values for $A$ and $B$ factors, we recalculate the position of the hyperfine components which are not observed. The calculated splitting of the HFS components $1\rightarrow{2}$ and $2\rightarrow{1}$ of the $\SI{402.0}{\nm}$ line is $\SI{18}{\MHz}$, explaining the unresolved resonances of these lines.
As a test of consistency, we calculate the complete shape of the saturation HFS spectra according to Ref.~\cite{Borde:1979}, fit it to the experimental signal and find that it confirms the results presented above. 

\begin{table}[b!th]
\centering
\caption{\label{table:splittingrates} Hyperfine splitting factors of $^{229}\Th^{+}$ electronic levels.  The second row for the ground state shows the values obtained in Ref.~\cite{Kalber:1989}.
$^{(a)}$These results represent the fit where the $A$ and $B$ parameters for the ground state are fixed to the values from Ref.~\cite{Kalber:1989}.}
\begin{ruledtabular}
\begin{tabular}{*{1}{c|}*{2}{c|}*{1}{c}}
Level $\left[\mathrm{cm}^{-1}\right]$    & $A$  $\left[\mathrm{MHz}\right]$     &  $B$ $\left[\mathrm{MHz}\right]$    & Reference  \\
\hline
0   & -444.2(3.4)        &  308(13)      &  This work \\
    & -444.2(1.9)        &  303(6)    & Ref.~\cite{Kalber:1989} \\
\hline
24874    & 489.2(3.7)    &  -409(18)    &  This work \\
   &488.9(3.7)    &  -413(16)   &  This work$^{(a)}$\\
\hline
25027 & -45(19)    &  &   This work                                    \\
  &-42(16)  &  & This work$^{(a)}$            \\
\end{tabular}
\end{ruledtabular}
\end{table}

The IS of $^{229} \Th^{+}$ is determined from the hyperfine centroid, corresponding to the position of the line for $A=B=0$. For the $\SI{402.0}{\nm}$ line, the centroid is $\approx$ \SI{810}{\MHz} red-detuned from the position of the $3\rightarrow{4}$ component.
The isotope shift of  $^{229} \Th^{+}$ for this line is \SI{18.86(7)}{\GHz}, in agreement with earlier, less precise measurements~\cite{Vernyi:1959, Vernyi:1960}. The uncertainty is determined by the precision of the frequency intervals measurements and the calculated value of the HFS centroid.  
The GS of $\Th^{+}$ yields a mixture of electronic configurations with leading components $6d^{2}7s$ and $6d7s^{2}$ while the $24874_{5/2}$ level contains 26\% $6d7s7p$ and 20\% $5f6d^{2}$ ~\cite{Zalubas:1974}. For this transition the FS of the ground state is dominant  ~\cite{Vernyi:1960} and determines the positive IS of the \SI{402.0}{\nm} line.

The $\SI{399.6}{\nm}$ line shows a negative IS of \SI{-3.67(5)}{\GHz}. The centroid of the line is shifted at $\approx$ \SI{1.5}{\GHz} to the lower frequency from the resonance $4\rightarrow{3}$ (see Fig.~\ref{fig:hfs_399}a). 
The probability to observe a negative IS in $\Th^{+}$ for a line originating from the GS was predicted to be very small ~\cite{Vernyi:1960}
since the $s$-electrons make the greatest contribution to the IS.
For the odd parity levels a negative IS is observed only for the transition from the GS to the $8379_{7/2}$ level which has the leading electron configuration $fs^{2}$.
According to ~\cite{Zalubas:1974} the leading configuration of the  $25027_{1/2}$ state is $5f6d^{2}$. The analysis in Ref. ~\cite{Stukenbroeker:1953} indicates a weak FS for the $5f6d^{2}$ electron configuration which means that the IS of the transition should be determined by the energy shift of the ground state which leads to a positive IS.
A potential explanation for the negative IS  would be a prior misidentification of the  $25027_{1/2}$ level and the presence of a large admixture of a $7s^2nl$ configuration. 

The field shift for isotope $A'$ with respect to isotope $A$ is defined as:
\begin{math} \Delta \omega^{A', A} =
K_\mathrm{FS}\, \delta \langle r^2 \rangle^{A', A}, \end{math}  where $K_\mathrm{FS}$ is the field shift constant and
\begin{math} \delta \langle r^2 \rangle^{A', A}= \langle r^2 \rangle^{A'} -
\langle r^2\rangle^{A}.\end{math}
The FS operator, $H_\mathrm{FS}$, is a one-particle operator
which modifies the Coulomb potential inside the nucleus. We use the
``finite field'' method, which means that a perturbation is added to
the initial Hamiltonian with the arbitrary coefficient $\lambda$:
$H\rightarrow H_\lambda= H + \lambda H_\mathrm{FS}$. We find eigenvalues
$E_\lambda$ by direct diagonalization of $H_\lambda$ and then find
the FS coefficient of an atomic level   as ~\cite{Korol:2007, Kozlov:2005}:
\begin{equation} 
 K_\mathrm{FS} = \frac{5}{6r_N^2} \frac{\partial
 E_\lambda}{\partial\lambda}\,
\label{eq1}
\end{equation}
where the nucleus is modeled as
a homogeneously charged ball of radius $r_N$.

\begin{table}
\caption{\label{Tab8}
Comparison of the g-factors and excitation energies (cm$^{-1}$) of Th$^+$ odd $J=1/2$ states calculated using CI+all-order method with experimental data compiled in Ref. \cite{Zalubas:1974} (see also: \cite{Data:th}). 
The last column gives finite size IS constants $K_{FS}$ in GHz/fm$^2$. All values are counted from the GS.}
\begin{ruledtabular}
\begin{tabular}{llc|cc|cc|c}
\multicolumn{2}{c}{State}&
\multicolumn{1}{c|}{$J$}&
\multicolumn{2}{c|}{g-factor}&
\multicolumn{2}{c|}{Energy}&
\multicolumn{1}{c}{$K_{FS}$}\\
\multicolumn{1}{c}{Ref.\cite{Zalubas:1974}}&
\multicolumn{1}{c}{Calc.}&
\multicolumn{1}{c|}{}&
\multicolumn{1}{c}{Calc.}&
\multicolumn{1}{c|}{Exp.}&
\multicolumn{1}{c}{Calc.}&
\multicolumn{1}{c|}{Exp.}&
\multicolumn{1}{c}{Calc.}\\
\hline
$                   5f6d7s              $                   & $                   5f6d7s              $                   &     1/2                 &       0.243               &                   0.255               & 12390               &                   11725               &     -82.1               \\
$                   5f6d7s              $                   & $                   5f6d7s              $                   &     1/2                 &       0.558               &                   0.523               & 14805               &                   14102               &     -96.7               \\
$                   5f6d7s              $                   & $                   5f6d7s              $                   &     1/2                 &       2.577               &                   2.565               & 16207               &                   15324               &     -65.9               \\
$                   5f6d7s              $                   & $                   5f6d^2              $                   &     1/2                 &       1.618               &                   1.080               & 18307               &                   17838               &     -120.4              \\
$                   5f6d^2              $                   & $                   5f6d^2              $                   &     1/2                 &       0.405               &                   1.007               & 19309               &                   18568               &     -126.4              \\
$                   5f6d7s              $                   & $                   5f6d^2+$            &                   1/2   &                   0.730   &                   0.770               &                   23456               & 22355               &                   -100.6              \\
$                                                                                 $                   &                   $     5f6d7s              $       &                                       &                                       &                     &                                       &                         &                           \\
$                   \bf{5f6d^2}         $                   & $                   \bf{7s^27p+}        $                   &     \bf{1/2}            &       \bf{0.568}          &                   \bf{0.600}          & \bf{25445}          &                   \bf{25027}          &     \bf{5.8}            \\
$                   $                   &                   $ \bf{6d7s7p}         $                   &                         &                   \       &                   \                   &                   \                   &                     &                                       \\
$                   5f6d^2              $                   & $                   5f6d^2              $                   &     1/2                 &       2.108               &                   0.725               & 26024               &                   25266               &     -119.7
                                        \end{tabular}
\end{ruledtabular}
\end{table}
 We use the CI+all-order method, with modified Hamiltonian given above, to carry out
  the field shift calculation. All computations are repeated three times: with $\lambda=0$ and $\lambda=\pm 0.01$.
  Then, Eq.~(\ref{eq1}) is used to evaluate the field shift constant. 
  The general description of the CI+all-order method and its  application to the calculation of energies, transition
  properties, and polarizabilities are given in \cite{Safronova:2009,Safronova:2014,Safronova:2014a}.
  We have modified the CI part of the method to be able to successfully generate much higher excited states than previously possible.
	This advancement will allow further applications of the CI+all-order method for
  highly-excited states and systems with dense spectra.
   Another issue of the application of the CI+all-order method to IS is the requirement of very high numerical accuracy since the correction
   due to IS is small.
   Therefore, we improved and  thoroughly tested the numerical accuracy of the approach such as convergence of various iteration procedures.

While only $ns$ orbitals have significant overlap with the nucleus, with $p_{1/2}$ having much smaller overlap, other states
will be affected by the inclusion of the IS effects due to the self-consistent nature of the potential. All of the $ns$ orbitals become slightly less bound, leading to the expansion of the
total self-consistent electronic  potential. This, in turn leads to changes of energies of the other states
as they become more deeply bound in the modified potential. The IS of $6d$ and $5f$ electrons will be with the
opposite sign of the $7s$ electron IS. One-electron estimates give the IS of the  $5f$ electron being about 60\% of the
$7s$ IS with an opposite sign.

The results of the CI+all-order  calculation
of the g-factors and excitation energies of Th$^+$ odd $J=1/2$ states are compared with experiment \cite{Zalubas:1974,Data:th} in Table~\ref{Tab8}. 
We find good agreement for the energy and g-factor of the  $25027_{1/2}$ level of interest.
The last column gives differential finite size IS constants $K_{FS}$ in GHz/fm$^2$ for $J=1/2$ levels, relative to the ground state.
The table clearly demonstrates the
opposite sign of the isotope shift constant for this level in comparison to the other states.
Our calculations show $7s^27p$ with  a contribution of 36\% and $6d7s7p$ with 34\% as the two dominant configurations for the $25027_{1/2}$ level, which explains the observations of the negative IS. Contributions from $6d^27p$ are 14\% and from $5f6d^2$ 4\% only.
Using the value  $\delta \langle r^2 \rangle^{229,232}=-0.33(5)$~fm$^2$ ~\cite{Kalber:1989,Angeli:2013} for the difference in the rms radii, we get \SI{-1.9}{\GHz} for the CI+all-order IS of the 399.6~nm line, in good agreement with the experimental value of \SI{-3.67(5)}{\GHz}.
The difficulty of this particular calculation is the very large correlation correction for the $5f$ electrons and
very strong configuration mixing for the $J=1/2$ states.  Our value for the \SI{402.0}{\nm} IS is 19.0~GHz, in excellent agreement with the experimental results of \SI{18.86(7)}{\GHz}.

We have also carried out calculations using CI+many-body-perturbation theory (MBPT) \cite{Dzuba:1996} previously employed for IS calculations \cite{BFK06}. The CI+MBPT results for the \SI{399.6}{\nm} and
\SI{402.0}{\nm} lines are 8.1~GHz and 26.0~GHz, respectively, in very poor agreement with experiment. This demonstrates the success of the CI+all-order method and the need for the inclusion of the higher-order effects.

In summary, we have demonstrated that the CI+all-order method can accurately describe isotope shifts in such complicated systems as Th$^+$. In turn, the IS measurements in actinides represent excellent benchmark tests for the development of these
highly accurate theoretical methodologies. The results of this work are important for the selection of efficient transitions for the excitation of the $^{229}$Th isomer, as the IS indicates the strength of the Coulomb interaction between electrons and nucleus that is relevant for NEET processes.  The comparison of isotope and isomer shifts will give access to the Coulomb contribution to the nuclear transition energy, resolving a dispute about largely different theoretical predictions of the sensitivity of the nuclear transition frequency to the value of the fine structure constant \cite{Berengut:2009}.

This work was supported in part by U. S. NSF Grant
No. PHY-1404156 and RFBR Grant No. 14-02-00241 and in part by EU FET-Open project 664732 - nuClock.


\end{document}